\documentclass[aip, pop, reprint, onecolumn]{revtex4-1}
\usepackage{natbib}

\usepackage{graphicx}
\usepackage{epstopdf}
\usepackage{amsmath}

\usepackage{amssymb}
\usepackage{color}
\usepackage{xspace}

\newcommand\eg{\textit{e.g.}\xspace}
\newcommand\ie{\textit{i.e.}\xspace}

\renewcommand\Im{\mbox{$\mathrm{Im}$}}

\newcommand{\vpar}{\ensuremath{v_{\parallel}}}
\newcommand{\vperp}{\ensuremath{v_{\perp}}}
\newcommand{\wpar}{\ensuremath{w_{\parallel}}}
\newcommand{\wperp}{\ensuremath{w_{\perp}}}
\newcommand{\xpar}{\ensuremath{x_{\parallel}}}
\newcommand{\xperp}{\ensuremath{x_{\perp}}}

\newcommand{\vd}{v_d}
\newcommand{\vb}{\ensuremath{\bar{v}}}

\newcommand{\vkt}[1]{\tilde{#1}}

\newcommand\vkg{\vkt{g}}

\def \e { \mbox{$\mathrm{e}$} }

\def \vth {\mbox{$v_{\mathrm{T}}$}}

\def \vs {v_*}

\begin{document}

\title{On the nonlinear stability of a quasi-two-dimensional drift kinetic model for ion temperature gradient turbulence}
\author{G. G. Plunk}
\email{gplunk@ipp.mpg.de}
\affiliation{Max Planck Institute for Plasma Physics, Wendelsteinstr. 1, 17491 Greifswald, Germany}

\begin{abstract}
We study a quasi-two-dimensional electrostatic drift kinetic system as a model for near-marginal ion temperature gradient (ITG) driven turbulence.  A proof is given of the nonlinear stability of this system under conditions of linear stability.  This proof is achieved using a transformation that diagonalizes the linear dynamics and also commutes with nonlinear $E\times B$ advection.  For the case when linear instability is present, a corollary is found that forbids nonlinear energy transfer between appropriately defined sets of stable and unstable modes.  It is speculated that this may explain the preservation of linear eigenmodes in nonlinear gyrokinetic simulations.  Based on this property, a dimensionally reduced ($\infty\times\infty \rightarrow 1$) system is derived that may be useful for understanding dynamics around the critical gradient of Dimits.
\end{abstract}

\maketitle

\paragraph{\bf Motivation.}  The most unstable ion temperature gradient (ITG) modes in toroidal fusion devices, such as tokamaks and stellarators, rely on a drift resonance due to curvature in the magnetic guide field.  One signature of this is a concentration of turbulent activity around regions of ``bad curvature.''  This reflects the linear instabilities themselves which peak or ``balloon'' in these regions.  For  sufficiently unstable modes, two conditions can be simultaneously met \cite{plunk-pop-itg-mode}: (1) the ballooning is strong (the mode is strongly localized) and (2) the timescale associated with ion propagation along this structure is long compared to the mode frequency (\ie the parallel transit frequency is small), rendering the dynamics nearly two-dimensional.  For these modes, the scale lengths perpendicular to the magnetic field are significantly larger than the ion Larmor radius so finite Larmor radius (FLR) effects can be considered small.  

If all of the conditions above are met, even if only for a range of spatial scales, it is appropriate to study the two-dimensional drift kinetic equation to understand the dynamics that results.  Furthermore, a precise understanding of the basic properties of this equation is, generally speaking, essential as a basis for the understanding of the general gyrokinetic system, of which it is a limiting case.

There are two important fundamental questions that further motivate this work.  First, why does plasma turbulence sometimes retain the features of the linear instabilities that drive it?  In the fully nonlinear state, one might expect the plasma fluctuations to become disordered to the extent that linear modes cannot be recognized by their spatial structure, frequency and growth rate.  Yet, all of these features have indeed been observed in fully developed turbulence \cite{dannert-jenko, goerler-jenko, hatch-terry}, even when it is strongly driven \cite{plunk-banon-navarro}.  Second, what determines whether turbulence will exist in the absence of linear instabilities?  Whenever there is non-uniformity of the background, there is a source of free energy to excite turbulence, and many plasma systems are known to support turbulence even when there are no linearly unstable modes to drive it.\cite{friedman, schek-highcock, landreman-plunk-dorland}  It should therefore be valuable to identify linearly stable cases in which turbulence provably does not exist, and to understand the underlying reasons.

\paragraph{\bf Basic analysis.} Let us consider the following two-dimensional drift-kinetic equation for the ion species of a magnetized plasma (obtained from the gyrokinetic equation by taking the Larmor radius to be small and ignoring parallel streaming):

\begin{equation}
\frac{\partial f}{\partial t} + i k_y \vd f + i k_y(\vd - \vs)\varphi F_0 = \displaystyle{\sum_{{\bf p},{\bf q}} } \epsilon_{{\bf k}{\bf p}{\bf q}} \varphi^*({\bf p}) f^*({\bf q}),\label{drift-kinetic-eqn}
\end{equation}

\noindent where $\varphi = q_i\phi/T_0$ is the normalized electrostatic potential, $F_0$ is the background ion distribution, and the perturbed distribution function is $f({\bf k}, \vperp, \vpar, t)$ where ${\bf k} = \hat{\bf x} k_x +\hat{\bf y} k_y$.  The nonlinear coupling coefficient is $\epsilon_{{\bf k}{\bf p}{\bf q}} = \delta({\bf k} + {\bf p} + {\bf q})T_0/(q_iB)\hat{\bf z}\cdot({\bf p}\times{\bf q})$ is antisymmetric under the exchange of any two components, \ie has the symmetry of the Levi-Civita tensor.  The electrostatic potential is determined by the quasi-neutrality constraint, assuming a modified Boltzmann electron response.  The quasi-neutrality constraint is written

\begin{equation}
(1-\delta_{k_y}) \tau \varphi = \frac{2\pi}{n_0}\int_{-\infty}^{\infty} d\vpar \int_{0}^{\infty}\vperp d\vperp f,\label{qn-eqn}
\end{equation}

\noindent where $\delta_{k_y}$ is the discrete delta function.  Note that this implies the zonal ion density is zero (at this order in the small Larmor radius expansion).  To close this system we need an equation for the zonal electrostatic potential ($k_y = 0$), which is given by the vorticity equation (obtained from the gyrokinetic system, including terms of order $k^2 \rho^2$; see \citet{plunk-njp} Eqn. D.9)

\begin{equation}
\frac{\partial}{\partial t} (k_x^2 \varphi) = \sum_{{\bf p},{\bf q}} \epsilon_{{\bf k}{\bf p}{\bf q}} \varphi^*({\bf p}) \left[q^2 \varphi^*({\bf q}) - {\bf p}\cdot{\bf q}T_{\perp}^*({\bf q})\right],\quad k_y = 0.\label{z-vorticity-eqn}
\end{equation}

\noindent We use the following additional conventions: $F_0 = n_0(\vth^2\pi)^{-3/2}\exp[-v^2/\vth^2]$, $v^2 = \vperp^2 + \vpar^2$, $\tau = T_{i0}/(ZT_{e0})$, $Z = q_i/|q_e|$, $T_0 = T_{i0}$,  $T_{\perp} = (T_0 n_0)^{-1}2\pi \int \vperp d\vperp d\vpar f (m \vperp^2/2)$, $\vd = \rho \sqrt{2}/(\vth R)[\vperp^2/2 + \vpar^2]$ and $\vs = \rho\vth/(L_n\sqrt{2}) (1 + \eta[v^2/\vth^2 - 3/2])$, $L_n^{-1} = d\ln n_0/dx$, $L_T^{-1} = d\ln T_0/dx$, $\eta = L_n/L_T$ .  

In the presence of a sufficiently large temperature gradient and magnetic curvature drift of the correct sign (\ie bad curvature), this system is linearly unstable to ITG modes of the ``toroidal branch.''  A stability analysis of this mode was done by \citet{biglari} who found a sufficient condition for stability, $L_T/L_n \geq 3/2$.  Although a necessary and sufficient stability criterion could not be obtained analytically, the quantity $\kappa = R/L_T$ is typically cited as the important instability parameter governing the toroidal ITG branch.  For the present purposes, we need not specify the value of the parameters in our system; we will simply specify whether or not the linear ITG mode is stable.  

Now there is a simple quadratic form that is a dynamical invariant of the system \ref{drift-kinetic-eqn}-\ref{qn-eqn}.  It is found by multiplying Eqn.~\ref{drift-kinetic-eqn} by the complex conjugate $f^*({\bf k})$ and then taking the real part of the result; this annihilates the $\vd f$ term.  Then, we divide  by the function $\kappa(\vperp, \vpar) = (\vd - \vs)F_0$, integrate over velocity, and use Eqn.~\ref{qn-eqn} to annihilate the term that goes as $i k_y f^* \varphi + \mathrm{c.c.}$.  Finally, we sum over ${\bf k}$, and use the antisymmetry property of the nonlinearity to show that its total contribution is zero.  The result is

\begin{equation}
\frac{d}{dt}\left[\sum_{\bf k} 2\pi \int \vperp d\vperp d\vpar  \frac{|f|^2}{\kappa}\right] = 0.\label{trivial-conserv-eqn}
\end{equation}

This is a useful constraint if the function $\kappa$ is sign definite, because then the integral defines a norm for the function $f$.  In this case Eqn.~\ref{trivial-conserv-eqn} is a statement of nonlinear stability.  The sign-definiteness of $\kappa$ can be guaranteed if the stability criterion $L_T/L_n \geq 3/2$ is satisfied, but this is not a very useful criterion as already mentioned.  In what follows nonlinear stability is proved in a much stronger sense, \ie by only assuming linearly stability it is shown that Eqn.~\ref{drift-kinetic-eqn} is also nonlinearly stable in the sense that an appropriately defined norm is invariant in time.

\paragraph{\bf Nonlinear stability.}  We now prove a nonlinear stability theorem by explicit use of the transformation that diagonalizes the linear operator.  The proof relies crucially on the existence of a complete set of eigenmodes of the linear operator, which we now assume to be stable.  Because the dependence of this operator on ${\bf k}$ is only in the overall multiplicative factor of $k_y$, it should be immediately clear that the eigenmode analysis is independent of ${\bf k}$.  Consequently, linear diagonalization commutes with the nonlinear $E\times B$ advection and the transformed system is manifestly stable.  Let us show this in detail.

First, we reduce the velocity space of system \ref{drift-kinetic-eqn} by using the change of variables $(\vperp, \vpar) \rightarrow (\vb, \lambda)$.  The idea is that the linear phase-mixing in Eqn.~\ref{drift-kinetic-eqn} is really one dimensional, so we should be able to eliminate the ignorable dimension.  The transformation can be done in two steps, defining first the polar coordinates $w = (\wperp^2 + \wpar^2)^{1/2}$ and $\theta$ and corresponding cartesian coordinates $\wperp = \vperp/\sqrt{2}$ and $\wpar = \vpar$, such that $\theta$ is the angle between the $\wperp$-axis and the vector $(\wperp, \wpar)$.  Then we define $\vb =  \vd =  w^2/c_d$ and $\lambda = \sin(\theta)$, where $c_d = \vth R/(\sqrt{2}\rho)$.  The equation may be integrated over $\lambda$ to obtain the reduced system in $g({\bf k},\vb, t) = n_0^{-1}\vb^{1/2}c_d^{1/2}\int_{-1}^{1} d\lambda f$:

\begin{equation}
\frac{\partial g}{\partial t} + i k_y \vb g + i k_y G(\vb) n = \displaystyle{\sum_{{\bf p},{\bf q}} } \epsilon_{{\bf k}{\bf p}{\bf q}} \varphi^*({\bf p}) g^*({\bf q}),\label{dk-reduced-eqn}
\end{equation}

\noindent where 

\begin{equation}
n = \int_0^{\infty} d\vb g(\vb),
\end{equation}

\noindent and

\begin{equation}
G(\vb) = \frac{2}{\tau}\sqrt{\frac{\vb c_d^3}{\pi}}\int_{-1}^{1} \frac{\kappa(\vb,\lambda)}{\vth^3} \exp\left[-\frac{c_d\vb(2-\lambda^2)}{\vth^2}\right] d\lambda.
\end{equation}

\noindent In the new variables we write $\kappa(\vb, \lambda) = \vb - c_*(1 + \eta((2 - \lambda^2)\vb c_d/\vth^2-3/2))]$ with $c_* = \rho\vth/(L_n\sqrt{2})$.  The linearized system has a continuum of neutrally stable solutions, and in the present form it is almost directly amenable to solution by the Morrison transform;\cite{morrison-pfirsch, morrison-aa} see also \citet{plunk-landau} for a recent application of this technique.  However Eqn.~\ref{dk-reduced-eqn} is different in form from the classical problem of Landau damping in that the velocity-space domain is only semi-infinite, $\vb \in [0,\infty)$.  The continuous spectrum is also only semi-infinite, and the required mathematical tools (Hilbert and Fourier transforms) are defined on an infinite domain.  To circumvent this difficulty, we can simply extend the space $\vb$ to the full domain.  Formally we consider the new system on $\vb \in (-\infty,\infty)$ for the function $g^{\mathrm{(e)}}(\vb)$, $n^{\mathrm{(e)}}(\vb) = \int_{-\infty}^{\infty} d\vb g^{\mathrm{(e)}}(\vb)$ and $G^{\mathrm{(e)}}(\vb) = G(\vb)\Theta(\vb)$, where $\Theta$ is a step function, $\Theta[\vb] = 1$ for $\vb \geq 0$ and otherwise zero.  The linear dispersion relation for the discrete spectrum is equivalent to that of \ref{dk-reduced-eqn}, and the nonlinear solutions of the original system are a subset of the solutions of the extended system.  In what follows we will omit the superscripts of the extended system to keep the notation simple.

Let us define the Morrison transform in the notation of \citet{plunk-landau}

\begin{equation}
\vkg = \frac{g_+}{D_+} + \frac{g_-}{D_-},\label{M-tranform-eqn}
\end{equation}

\noindent where $D_\pm = 1 \pm 2\pi i G_\pm$.  The positive- and negative-frequency parts of an arbitrary function $g(v)$ are defined in terms of the Fourier transform by

\begin{equation}
g_{\pm}(u) = \pm \int_0^{\pm \infty} d\nu \e^{i \nu u} \int_{-\infty}^{\infty} dv\frac{\e^{-i \nu v}}{2 \pi} g(v).
\end{equation}

The Morrison transform can be directly applied to Eqn~\ref{dk-reduced-eqn} to obtain

\begin{equation}
\frac{\partial \vkg}{\partial t} + i k_y u \vkg = \displaystyle{\sum_{{\bf p},{\bf q}} } \epsilon_{{\bf k}{\bf p}{\bf q}} \varphi^*({\bf p}) \vkg^*({\bf q}).\label{dk-trans-eqn}
\end{equation}

\noindent The following conservation law is then obtained, completing our proof of nonlinear stability:

\begin{equation}
\frac{d}{dt}\sum_{\bf k}|\vkg|^2 = 0\label{nl-stable-eqn}
\end{equation}

Note that this applies point-by-point in velocity space $u$, in contrast to the quantity in Eqn.~\ref{trivial-conserv-eqn}.  Eqn.~\ref{nl-stable-eqn} constitutes a statement of stability because Eqn.~\ref{M-tranform-eqn} is an invertible transformation.  Invertibility is equivalent to completeness of the eigenmodes, which, as explained by \citet{vk-book}, is guaranteed if the function $D_+(u)$ has no zeros in the upper half plane.  This is true if and only if the dispersion relation for the linear system has no roots in the upper half plane.

\paragraph{\bf Unstable modes and ``energetic isolation.''}  It turns out that the above analysis may be extended to cases with a discrete spectrum, resulting in a theorem about energy flow in the unstable system.  Following the work of \citet{case}, we may use the orthogonality of the eigenmodes to separate the continuous and discrete parts of the eigenspectrum.  Numbering the discrete modes $1, 2, 3, ... , m$, we define the projection operator for each mode

\begin{equation}
{\mathcal P}_i \left[g\right] = \frac{g_i(u)}{C_i} \int d\vb h_i(\vb) g(\vb),
\end{equation}

\noindent where $g_i(u) = -G(u)/(u - u_i)$ is an eigenmode of Eqn.~\ref{dk-reduced-eqn}, $h_i(\vb) = -1/(\vb - u_i)$ is the corresponding eigenmode of an adjoint equation, and $C_i = \int_0^{\infty} dv G(v)/(v - u_i)^2$; see \citet{case} for details.  Note that the discrete spectrum generally includes pairs of damped and growing modes, and also stable non-singular eigenmodes (exceptional cases).  The projection operator may be applied to Eqn.~\ref{dk-reduced-eqn} to obtain a system that describes the linear evolution and nonlinear interaction among the sub-population of modes $\{g_i({\bf k}_1), g_i({\bf k}_2), ...\}$, indexed by the wavenumbers of the system.  That is, writing $g({\bf k}, \vb) = \int A(\nu, {\bf k})g_{\nu}(\vb) d\nu + \sum_i a_i({\bf k}) g_i(\vb)$ we have

\begin{equation}
\frac{\partial a_i({\bf k})}{\partial t} + i k_y u_i a_i({\bf k}) = \displaystyle{\sum_{{\bf p},{\bf q}} } \epsilon_{{\bf k}{\bf p}{\bf q}} \varphi^*({\bf p}) a_i^*({\bf q}),\label{ai-eqn}
\end{equation}

\noindent implying

\begin{equation}
\frac{d}{dt}\sum_{\bf k}\frac{|a_i|^2}{2} = \sum_{\bf k} k_y \Im[u_i] |a_i|^2.
\end{equation}

We may also isolate the continuous spectrum by simply applying the operator ${\mathcal P}' = 1 - \sum_{i} \mathcal{P}_i$ to Eqn.~\ref{dk-reduced-eqn}.  This yields an equation for $g' \equiv {\mathcal P}'[g]$ (of the same form of Eqn.~\ref{dk-reduced-eqn}) that only has a continuous stable linear spectrum.  To this system the invertible Morrison transform may be applied as done in the previous section.  The sum of the energy associated with these modes is an exact invariant of the full nonlinear equation, \ie Eqn.~\ref{nl-stable-eqn} for $\vkg'$.  In summary, we may decompose the system into sets of modes (\ie the sets $\{g_i({\bf k}_1), g_i({\bf k}_2), ...\}$ and $\{g_\nu({\bf k}_1), g_\nu({\bf k}_2), ...\}$) that can be dynamically coupled via the nonlinearity but are energetically de-coupled.  Note that for a set of unstable modes, the zonal components $a_i(k_x, k_y = 0)$ are linearly stable, but can be excited nonlinearly.  This energetic isolation of linear mode populations is an intuitive property: energy is passed between equivalent modes because (1) the linear eigenmodes do not depend on the wavenumber ${\bf k}$ and (2) nonlinear mode coupling preserves the velocity dependence of $g$.

Note that a consequence of energetic isolation is that damped eigenmodes \cite{hatch-terry} cannot be excited by nonlinear interaction with the set of unstable eigenmodes.  This does not contradict studies of fluid and gyrokinetic systems \cite{makwana-pop, makwana-prl} that demonstrate a significant degree of such excitation, because those systems have additional mechanisms, e.g. nonlinear phase mixing, that enable energetic exchange between different mode types.  However, the present work does imply that such energy flows should be small to a degree determined by how well the assumptions of our system (namely $k^2\rho^2 \ll 1$ and small collisional and parallel transit frequencies) are satisfied.

\paragraph{\bf Dimensional reduction}  The exact energetic isolation of eigenmodes suggests that a dramatic simplification could be attempted.  Let us assume the stable modes can be neglected (in a real plasma, an additional mechanisms like collisional dissipation could drain the energy of the stable spectrum) and that there is precisely one unstable mode for each wavenumber ${\bf k}$ with $k_y \neq 0$, which, dropping the indices, we notate as $a$.  Then by Eqn.~\ref{qn-eqn}, the non-zonal part of the electrostatic potential is proportional to the amplitude of the unstable mode, \ie $\varphi({\bf k}) = 2\pi (c_d/\tau) a({\bf k})$ for $k_y \neq 0$.  Now we can divide the nonlinear coupling terms on the right hand side of Eqn.~\ref{ai-eqn} into three groups, (1) the non-zonal/non-zonal interaction terms, $p_y \neq 0$ and $q_y \neq 0$, (2) interactions involving the zonal potential, $p_y = 0$ and $q_y \neq 0$, and (3) interactions involving the zonal part of $a$, $p_y \neq 0$ and $q_y = 0$.  The first group of interactions clearly sum to zero by the antisymmetry of $\epsilon$.  The final group must also be zero because $a({\bf k}) = 0$ for $k_y = 0$, which follows from Eqn.~\ref{qn-eqn} and the fact that the eigenmodes have unit density.  What remains is a very simple system consisting of two equations.  The first is

\begin{equation}
\frac{\partial a({\bf k})}{\partial t} + i k_y u_i a({\bf k}) = \sum_{p_y = 0, q_y \neq 0} \epsilon_{{\bf k}{\bf p}{\bf q}} \varphi^*({\bf p}) a^*({\bf q}),\label{a-eqn}
\end{equation}

\noindent where the sum is performed over the set of ${\bf p}$ and ${\bf q}$ satisfying $p_y = 0$ and $q_y \neq 0$ .  This equation is then closed by Eqn.~\ref{z-vorticity-eqn} for the zonal vorticity, which can now be written in terms of the drift wave amplitudes:

\begin{equation}
\frac{\partial}{\partial t} (k_x^2 \varphi) = \sum_{{\bf p},{\bf q}} \epsilon_{{\bf k}{\bf p}{\bf q}} a^*({\bf p}) a^*({\bf q})\left[\alpha^2q^2  - \alpha \beta {\bf p}\cdot{\bf q}\right],\quad k_y = 0.\label{z-vorticity-a-eqn}
\end{equation}

\noindent where $\alpha = 2\pi c_d/\tau$ and 

\begin{equation}
\beta = \frac{2\alpha}{\sqrt{\pi}}\int_0^{\infty}\xperp^3 d\xperp\int_{-\infty}^{\infty} d\xpar \left(\frac{\vd - \vs}{\vd - u_i}\right)\exp(-\xperp^2 -\xpar^2).
\end{equation}

Eqn.~\ref{a-eqn} describes the nonlinear zonal shearing of a set of unstable drift waves, and the voriticity equation \ref{z-vorticity-a-eqn} describes the nonlinear response of the zonal flows to the bath of drift-waves.  Note that some kind of dissipation is needed to make this system well-behaved since there is no mechanism to saturate the unstable modes: if there are finite amplitude drift waves $a({\bf k})$, the set will always grow in energy.  If small-scale dissipation were to be added, then zonal shearing would be able to transfer energy to large $k_x$ sink and presumably cause saturation.  Note that the non-density components of the zonal modes become larger in amplitude, and dynamically important above the nonlinear critical gradient, and are responsible for nonlinear zonal flow decay via the tertiary instability \cite{rogers-prl}.  Thus it seems likely that the condition $a({\bf k}) = 0$ for $k_y = 0$, restricts the regime of validity of this system to around or below the nonlinear critical gradient of Dimits \cite{dimits}.

\paragraph{\bf Conclusion.}  We have shown that the two-dimensional drift kinetic system, Eqns.~\ref{drift-kinetic-eqn}-\ref{z-vorticity-eqn}, satisfies two properties: (1) it is nonlinearly stable when linearly stable and (2) it generally only allows equivalent linear eigenmodes to exchange energy nonlinearly.  The first property means that subcritical turbulence cannot be excited unless by a mechanism not present in our system, \eg by three-dimensional mode coupling \cite{friedman} or the parallel velocity gradient drive \cite{schek-highcock}.  This is an encouraging fact because it seems to limit the contribution of the magnetic drift resonance as a turbulent drive to only cases where it can induce a linearly unstable mode.  The second property, \ie the fact that eigenmode sets are energetically isolated, may help explain observations \cite{hatch-prl} that strongly-driven ITG turbulence preferentially dissipates energy by cascade to smaller scales, rather than by nonlinear transfer from an unstable mode to a damped mode at a similar scale.  Indeed, as the turbulence is more strongly driven, it peaks at larger scales, $k\rho \ll 1$, making our drift kinetic system more applicable.  Energetic isolation may also help explain why gyrokinetic turbulence retains recognizable features of the linear eigenmodes, even when strongly driven.  The preservation of linear modes is a mysterious but well-known property \cite{dannert-jenko, goerler-jenko, hatch-terry} of gyrokinetic turbulence with no accepted explanation.  Crucially, we have not assume weakness of the nonlinearity, so this result constitutes a new kind of mechanism for preserving linearity that might motivate quasilinear models when the assumption of weak nonlinearity is unjustified.

\bibliography{NL-stability-of-ITG}
\end{document}